\documentclass[
  reprint,
  superscriptaddress,
  amsmath,amssymb,
  longbibliography,
  aps,
  prl
]{revtex4-2}

\usepackage{bm}
\usepackage{booktabs}
\usepackage{graphicx}
\usepackage[caption=false]{subfig}
\usepackage{xcolor}
\usepackage[colorlinks=true,citecolor=blue,urlcolor=blue,linkcolor=blue]{hyperref}

\newcommand{\Mcut}{M_{\mathrm{cut}}}
\newcommand{\Mlow}{M_{\mathrm{low}}}
\newcommand{\Mhigh}{M_{\mathrm{high}}}

\begin{document}

\title{Soft decoding for quantum LDPC codes with experimental validation}

\author{Arda Aydin}
\affiliation{IonQ Inc.}
\affiliation{Department of ECE and Institute for Systems Research, University of Maryland, College Park, MD 20742}

\author{Edwin Tham}
\affiliation{IonQ Inc.}

\author{Nicolas Delfosse}
\affiliation{IonQ Inc.}

\author{Min Ye}
\affiliation{IonQ Inc.}

\date{\today}

\begin{abstract}
The decoder is a critical component of a fault-tolerant quantum computer, computing corrections based on parity-check measurements performed throughout the computation. A soft decoder supplements its output with a confidence score which, when used alongside post-selection, can substantially improve logical performance. We introduce a soft beam search decoder for quantum low-density parity-check codes that uses internal decoder data as a confidence metric, removing the need for extra computation. We perform circuit-level simulations of five quantum LDPC codes relevant for superconducting and trapped ion architectures equipped with our global soft decoder and we obtain up to $580\times$ logical-error suppression at a physical error rate of $10^{-3}$ while rejecting only $0.1\%$ of shots. Then, we simulate an error detected measurement, which is a core logical operation of the walking cat architecture, using a streaming version of soft beam decoder and we achieve up to $210\times$ error suppression while increasing the rejection probability by only $0.5$ percentage points. Finally, we revisit recent quantum LDPC code memory experiments on trapped ions, demonstrating that our soft decoder doubles the logical qubit lifetimes at the price of a mean rejection rate of $2.6\%$--$5.6\%$ per syndrome round, bringing all five codes into the beyond-breakeven regime.
\end{abstract}

\maketitle

\section{Introduction}

Soft decoding supplements a decoding result with information about its reliability. This information enables post-selection: low-confidence shots, quantum states or sub-circuits can be discarded to reduce the logical error rate of the accepted ensemble. Post-selection has long played a role in fault-tolerant quantum computation to perform error correction using post-selected cat states~\cite{shor1996fault}, logical zero and plus states~\cite{steane1997active, paetznick2011fault} or logical Bell states~\cite{knill2005scalable}, achieving record accuracy threshold by concatenation~\cite{knill2005quantum, aliferis2007accuracy}, to perform logical operations using magic states~\cite{bravyi2005universal, bravyi2012magic, li2015magic, gidney2024magic, bombin2024fault} or stabilizer states~\cite{gottesman1999quantum, zheng2018efficient, delfosse2025low, sunami2026entanglement, webster2026fast} and in quantum error mitigation~\cite{li2017efficient, temme2017error, cai2023quantum}. Its experimental utility has been demonstrated in quantum error correction~\cite{linke2017fault, marques2022logical, chen2022calibrated, bluvstein2024logical, paetznick2024demonstration, paetznick2026improved}, logical Clifford gates~\cite{van2023single, tham2025optimized, martiel2025low, brown2026mid} and magic-state preparation~\cite{gupta2024encoding, ye2023logical, pogorelov2025experimental, daguerre2025experimental, lacroix2025scaling, dasu2025breaking, sales2025experimental, rosenfeld2025magic}.

Confidence measures range from simple syndrome density criteria~\cite{english2025thresholds} to logical gaps that quantify ambiguity between competing logical classes~\cite{hutter2014efficient,bombin2024fault,smith2024mitigating}. Decoder confidence also supports error mitigation in logical circuits~\cite{dincua2025error,zhou2025error}, concatenated decoding~\cite{gidney2025yoked}, adaptive decoder switching~\cite{toshio2025decoder}, and faster lattice surgery~\cite{akahoshi2026runtime}. Resource states can also be post-selected before they are used in a computation, even when some syndrome information becomes available only afterward~\cite{staples2026scalable}. These applications call for confidence measures that identify unreliable outcomes with little computational overhead.

For quantum low-density parity-check (qLDPC) codes, the beam search decoder~\cite{ye2026beam} combines belief propagation (BP) with a bounded search over candidate error configurations, providing a tunable tradeoff between accuracy and runtime. It has been used to decode trapped-ion memory experiments~\cite{tham2026breakeven} and simulate coherent errors~\cite{hines2026simulating}, while streaming implementations support the walking cat architecture~\cite{tripier2026fault} and large-scale real-time decoding on a single CPU~\cite{ye2026real}. Subsequent work has developed layered beam search variants~\cite{pradhan2026fast} and used the beam search decoder as a numerical baseline for circuit-level benchmarks~\cite{leverrier2026approximating}. A related multistage rewinding decoder combines beam search with guided restarts of min-sum decoding and is benchmarked under code-capacity noise~\cite{taghipour2026multistage}.

Existing soft-output decoders differ in their applicability and computational cost. For surface codes, confidence can be estimated from the clusters produced by minimum-weight perfect matching and union-find decoders~\cite{meister2024efficient,kishi2026even}. Lee et al.~\cite{lee2026efficient} extend cluster-based confidence metrics to general qLDPC codes using cluster sizes and log-likelihood ratios. These metrics require clustering information and therefore do not directly apply to the beam search decoder. Other approaches obtain confidence from neural decoders~\cite{bausch2024learning,gu2026scalable,dentelski2026neural} or separately trained syndrome classifiers~\cite{haug2026machine}. A complementary approach uses repeated decoding: the two-round logical-error criterion (2R-LEC) of Xie et al.~\cite{xie2026simple} tests agreement between logical predictions under original and reweighted priors, while forced-gap post-selection~\cite{wills2026forced} uses a baseline decode and one additional constrained decode per logical observable. These methods provide effective post-selection for qLDPC codes at the cost of additional decoder invocations.

\begin{table*}[t]
\centering
\caption{Performance of the global soft beam search decoders for a $X$-basis memory experiment with $d$ rounds of syndrome extraction under circuit-level noise with noise rate $p=10^{-3}$: logical error suppression and mean BP iterations at target rejection rates.
Our hybrid approach achieves a better logical error rate than 2R-LEC~\cite{xie2026simple} using fewer BP iterations.
}
\label{tab:global_soft_comparison}
\small
\setlength{\tabcolsep}{3pt}
\renewcommand{\arraystretch}{1.0}
\begin{tabular*}{\textwidth}{@{\extracolsep{\fill}}ll*{8}{c}@{}}
\toprule
 & & & & \multicolumn{3}{c}{LER suppression}
         & \multicolumn{3}{c}{Mean BP iterations} \\
\cmidrule(lr){5-7}\cmidrule(lr){8-10}
Code & $[[n,k,d]]$ & Noise Model & Rejection Rate
 & Cutoff & Hybrid & 2R-LEC & Cutoff & Hybrid & 2R-LEC \\
\midrule
BB90 & $[[90,8,10]]$ & Uniform & $10^{-4}$
 & $20.0\times$ & $36.6\times$ & $8.33\times$
 & $6.414$ & $6.575$ & $13.708$ \\
BB144 & $[[144,12,12]]$ & Uniform & $10^{-4}$
 & $50.6\times$ & $69.9\times$ & $51.1\times$
 & $9.131$ & $9.170$ & $27.905$ \\
\midrule
Q70 & $[[70,6,9]]$ & Walking cat~\cite{tripier2026fault} & $10^{-4}$
 & $10.8\times$ & $48.9\times$ & $15.5\times$
 & $5.980$ & $7.592$ & $15.031$ \\
Q102 & $[[102,22,9]]$ & Walking cat~\cite{tripier2026fault} & $10^{-4}$
 & $3.1\times$ & $23.0\times$ & $17.1\times$
 & $9.381$ & $15.548$ & $23.756$ \\
Q54 & $[[54,2,10]]$ & Walking cat~\cite{tripier2026fault} & $10^{-3}$
 & $346\times$ & $580\times$ & $6.75\times$
 & $8.880$ & $8.996$ & $28.334$ \\
\bottomrule
\end{tabular*}
\end{table*}

Here we introduce a soft beam search decoder whose confidence metric is obtained directly from its ordinary execution. Our central observation is that the cumulative number of BP iterations executed across the visited search paths before convergence provides an effective measure of decoding reliability. Rejecting shots whose cumulative BP iteration count exceeds a chosen cutoff requires no additional decoder invocation, no explicit likelihood calculation, and no trained model. Enforcing this cutoff during decoding also bounds the worst-case BP work. We further introduce a selective hybrid that applies 2R-LEC only to shots in an intermediate convergence range.

Using global beam search decoding under circuit-level noise at $p=10^{-3}$, we demonstrate that iteration-cutoff post-selection achieves logical-error suppression of $124\times$ and $78\times$ for the $[[90,8,10]]$ and $[[144,12,12]]$ bivariate bicycle codes, respectively, while rejecting only $0.1\%$ of shots. For the Q54 code from~\cite{tripier2026fault}, the hybrid rule achieves $580\times$ suppression at the same physical error rate and rejection rate. Across all five tested codes, the hybrid rule achieves the largest logical error suppression among the soft decoding algorithms compared in Table~\ref{tab:global_soft_comparison}, while using only $32\%$--$65\%$ of the mean BP iterations required by 2R-LEC.

We also apply the iteration-cutoff rule to sliding-window decoding for real-time quantum error correction. In simulations of logical measurements with the Q54 and Q70 codes~\cite{tripier2026fault} at $p=10^{-3}$, we achieve up to $210\times$ logical-error suppression while increasing the rejection rate by only $0.5$ percentage points.

Finally, we reanalyze experimental qLDPC memory data from an IonQ trapped-ion quantum computer~\cite{tham2026breakeven}, using the same beam search configuration as the original hard-decoding analysis. Under hard decoding, these memories had logical lifetimes comparable to the measured physical-qubit lifetime. Across all five tested BB5 and GB4 codes, iteration-cutoff post-selection more than doubles the fitted logical lifetime at mean rejection rates of $2.6\%$--$5.6\%$ per syndrome round, bringing all five codes into the beyond-breakeven regime. These gains require neither additional quantum experiments nor an extra decoding pass to estimate confidence.

\begin{figure*}[t]
    \centering
    \subfloat{
        \includegraphics[width=0.48\textwidth]{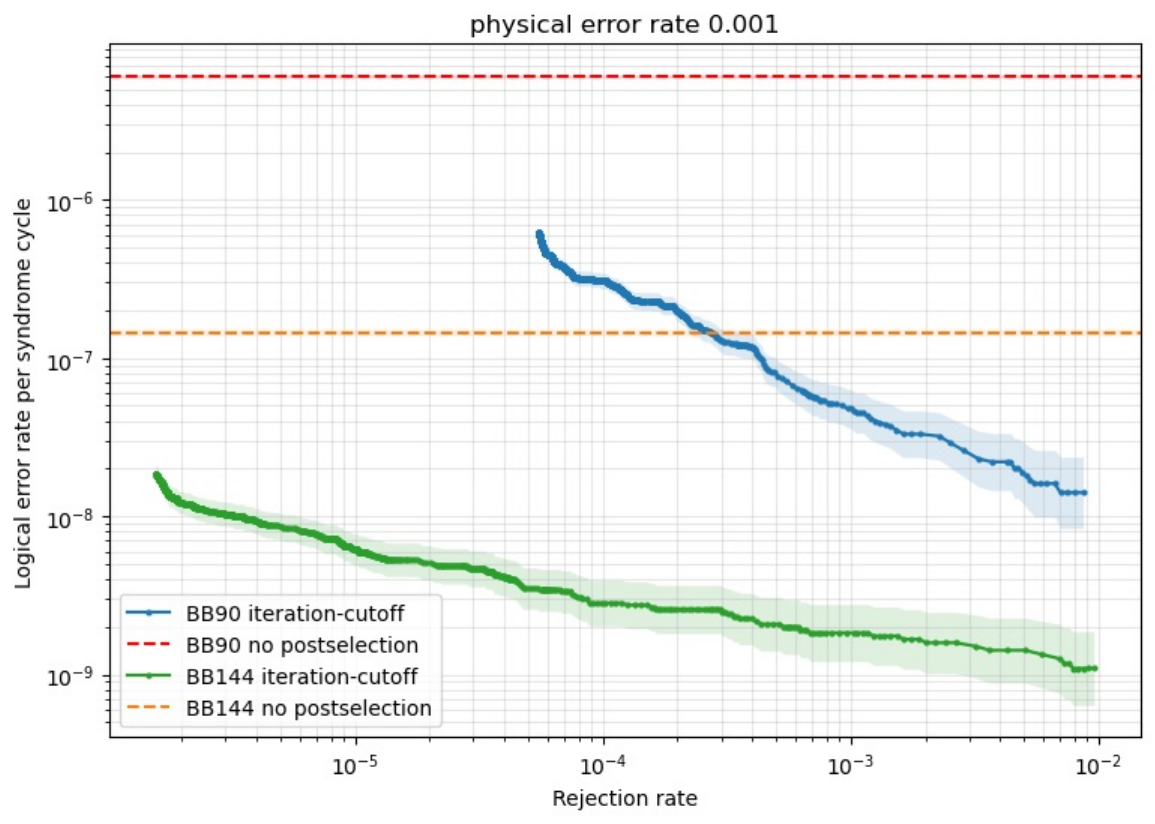}
    }
    \hfill
    \subfloat{
        \includegraphics[width=0.48\textwidth]{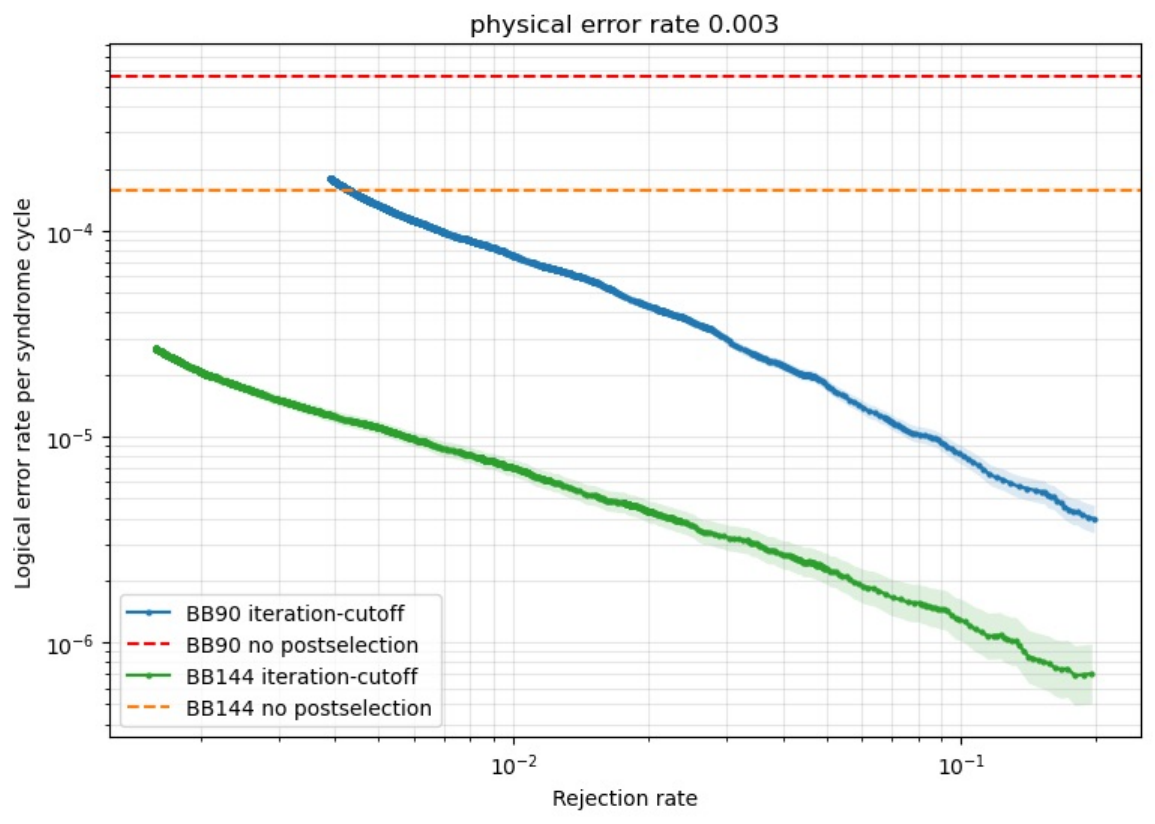}
    }
    \caption{Comparison of iteration-cutoff post-selection with the no-post-selection baseline for BB90 and BB144 at (left) $p=10^{-3}$ and (right) $p=3\times 10^{-3}$.}
    \label{fig:BB_90_144}
\end{figure*}

\section{Soft information from beam search convergence}
\label{sect:soft_info}

We first review the beam search decoder introduced in~\cite{ye2026beam}. The single-result beam search decoder is controlled by four parameters: \texttt{initial\_iters}, \texttt{iters\_per\_round}, \texttt{max\_rounds}, and \texttt{beam\_width}. For compactness, we denote these parameters by $I_0$, $I$, $R$, and $B$, respectively. The decoder begins with up to $I_0$ iterations of standard BP applied to a single unconstrained path; we refer to this initial stage as round~$0$. If BP finds a valid correction, namely, one satisfying the syndrome equations, the decoder terminates and immediately returns that correction. Otherwise, it performs up to $R$ rounds of masked BP, with at most $I$ BP iterations applied to each path. To construct round~$r$, the decoder selects the least reliable unfrozen variable of each path retained after round~$r-1$ and creates two child paths by fixing this variable to $0$ and $1$, respectively. Consequently, each path processed in round~$r$ contains $r$ frozen variables. The only distinction between masked BP and standard BP is that variables frozen along the current path are excluded from message passing on the Tanner graph.

To specify the order in which paths are decoded within each round, let $t_r=\min\!\left(2^r,B\right)$ denote the number of paths retained at the end of round~$r$, and label these paths $\pi^{(r)}_1,\ldots,\pi^{(r)}_{t_r}$ in decreasing order of reliability. For each retained path $\pi^{(r)}_i$, denote the two children described above by $\widetilde{\pi}^{(r+1)}_{i,0}$ and $\widetilde{\pi}^{(r+1)}_{i,1}$, where the second subscript indicates the value assigned to the newly frozen variable. Round~$r+1$ therefore contains $2t_r=\min\!\left(2^{r+1},2B\right)$ candidate paths. These candidates are decoded in the reliability order of their parent paths, with the $0$ child preceding the $1$ child for each parent:
\begin{equation}
    \widetilde{\pi}^{(r+1)}_{1,0},\,
    \widetilde{\pi}^{(r+1)}_{1,1},\,
    \widetilde{\pi}^{(r+1)}_{2,0},\,
    \widetilde{\pi}^{(r+1)}_{2,1},\,
    \ldots,\,
    \widetilde{\pi}^{(r+1)}_{t_r,0},\,
    \widetilde{\pi}^{(r+1)}_{t_r,1}.
    \label{eq:path-order}
\end{equation}

The single-result decoder terminates as soon as any candidate yields a valid correction. If none of the candidates in round~$r+1$ yields a valid correction, the decoder ranks all $2t_r$ candidates by reliability, retains the $t_{r+1}=\min\!\left(2^{r+1},B\right)$ most reliable candidates, and relabels them as $\pi^{(r+1)}_1,\ldots,\pi^{(r+1)}_{t_{r+1}}$ in decreasing order of reliability. If the search continues, these retained paths serve as the parent paths for the next round. Together with the initial BP run, Eq.~\eqref{eq:path-order} determines the complete sequence of BP iterations executed across all candidate paths visited before the decoder terminates.

For a syndrome vector $\boldsymbol{s}$ for which the decoder converges, we define its \emph{convergence iteration} $M(\boldsymbol{s})$ as the position in this sequence at which the first valid correction is found. If the initial BP run converges at its $\ell$th iteration, then $M(\boldsymbol{s})=\ell$. Otherwise, suppose that the first valid correction is found at the $\ell$th BP iteration of candidate $\widetilde{\pi}^{(r)}_{i,a}$ in round~$r$, where $1\leq \ell\leq I$, $1\leq i\leq t_{r-1}$, and $a\in\{0,1\}$. The ordering in Eq.~\eqref{eq:path-order} then gives
$$
M(\boldsymbol{s}) = I_0 + I\left( 2\sum_{j=0}^{r-2} t_j +2(i-1)+a \right) +\ell.
$$
For example, the \texttt{beam8\_230iters} configuration in~\cite{ye2026beam} uses $(I_0,I,R,B)=(30,20,10,8)$. The number $230$ in its name equals $I_0+RI$, the maximum number of BP iterations along a single root-to-leaf path. Since $M(\boldsymbol{s})$ accumulates the iterations executed across all visited paths, its maximum possible value is $M_{\max} = I_0+2I\sum_{j=0}^{R-1}t_j = 30+20\left(2+4+8+7\times16\right) = 2550$. If the first valid correction is found at the tenth BP iteration of $\widetilde{\pi}^{(4)}_{3,1}$, for instance, then $M(\boldsymbol{s})=30+20(2+4+8+5)+10=420$. If instead it is found at the seventh BP iteration of $\widetilde{\pi}^{(2)}_{2,0}$, then $M(\boldsymbol{s})=30+20(2+2)+7=117$.

\subsection{Post-selection rules}

We propose two post-selection rules for the beam search decoder. The first uses only the convergence iteration, whereas the second combines the convergence iteration with the 2R-LEC method from~\cite{xie2026simple}.

We first consider the iteration-cutoff rule. Among shots for which the beam search decoder converges, earlier convergence is empirically associated with a more reliable decoding result. We therefore use $M(\boldsymbol{s})$ as a confidence metric and accept a shot only if
$$
M(\boldsymbol{s})\leq \Mcut,
$$
where $\Mcut$ is a tunable cutoff threshold. Shots for which the decoder does not converge are always rejected. Varying $\Mcut$ produces a tradeoff between the rejection rate and the post-selected logical error rate.

This rule requires neither an additional decoder invocation nor an explicit likelihood calculation, because $M(\boldsymbol{s})$ is obtained directly from the execution of the underlying single-result decoder. Moreover, when the cutoff is enforced during decoding, a shot is rejected as soon as the cumulative iteration count reaches $\Mcut$ without a valid correction being found. The total number of BP iterations per shot is then upper bounded by $\Mcut$, reducing the worst-case decoding time.

The rule extends naturally to sliding-window decoding. Let $M^{(w)}(\boldsymbol{s})$ denote the cumulative number of BP iterations in the decoding of window~$w$. A shot is accepted only if every window required for that shot converges and $\max_w M^{(w)}(\boldsymbol{s})\leq \Mcut$.

We benchmark the iteration-cutoff rule on five code instances: Q54, Q70, and Q102 from~\cite{tripier2026fault}, and the $[[90,8,10]]$ and $[[144,12,12]]$ BB codes from~\cite{bravyi2024high}. We henceforth refer to the latter two codes as BB90 and BB144, respectively. As shown in Sec.~\ref{sect:simulation_results}, the iteration-cutoff rule achieves a favorable rejection--logical-error tradeoff for four of the five code instances but is less effective for Q102.

To improve post-selection for codes where the iteration-cutoff rule alone provides limited error suppression, we introduce a hybrid rule that combines the convergence iteration with the two-round logical-error criterion (2R-LEC) from~\cite{xie2026simple}. Since 2R-LEC also serves as a baseline in Section~\ref{sect:simulation_results}, we first briefly review the method. Let $\boldsymbol{p}=(p_1,\ldots,p_n)$ denote the original prior-probability vector, where $p_i=\Pr(e_i=1)$ is the prior probability that the $i$th binary error variable $e_i$ equals $1$. The first decoder pass returns a correction $\boldsymbol{c}_0$ and a logical prediction $\boldsymbol{\lambda}_0$. The second-pass prior vector $\boldsymbol{p}'=(p'_1,\ldots,p'_n)$ is then defined component-wise by
\begin{equation} \label{eq:2R-LEC}
    p_i' =
    \begin{cases}
        p_i^{\,b},
            & i\in\operatorname{supp}(\boldsymbol{c}_0),\\
        p_i,
            & i\notin\operatorname{supp}(\boldsymbol{c}_0),
    \end{cases}
    \qquad b>1.
\end{equation}
Since $0<p_i<1$, this transformation suppresses the prior probabilities of the variables in the support of the first-pass correction. The second pass decodes the same syndrome using $\boldsymbol{p}'$. If the second pass fails to converge, the shot is rejected. Otherwise, letting $\boldsymbol{\lambda}_1$ denote its logical prediction, 2R-LEC accepts the shot if and only if $\boldsymbol{\lambda}_1=\boldsymbol{\lambda}_0$.

Our hybrid rule introduces two convergence boundaries, $\Mlow<\Mhigh$, and acts on the result of the first pass according to
$$
    \begin{array}{lcl}
        M(\boldsymbol{s})\leq\Mlow
        &:& \text{accept after the first pass},\\
        \Mlow<M(\boldsymbol{s})<\Mhigh
        &:& \text{apply 2R-LEC},\\
        M(\boldsymbol{s})\geq\Mhigh
        &:& \text{reject after the first pass}.
    \end{array}
$$
A shot whose first pass does not converge is rejected. In the intermediate region, the shot is accepted if and only if the two passes produce the same logical prediction.

Let $f_{\mathrm{mid}}$ denote the fraction of all shots that fall in the intermediate region. Since only these shots require a second decoder pass, the mean number of decoder invocations is $1+f_{\mathrm{mid}}\leq 2$. By comparison, applying 2R-LEC to every shot requires two decoder invocations, while the forced-gap method of~\cite{wills2026forced} requires $K+1$ invocations for codes with $K$ logical observables.

\begin{figure*}[t]
    \centering
    \subfloat{
        \includegraphics[width=0.48\textwidth]{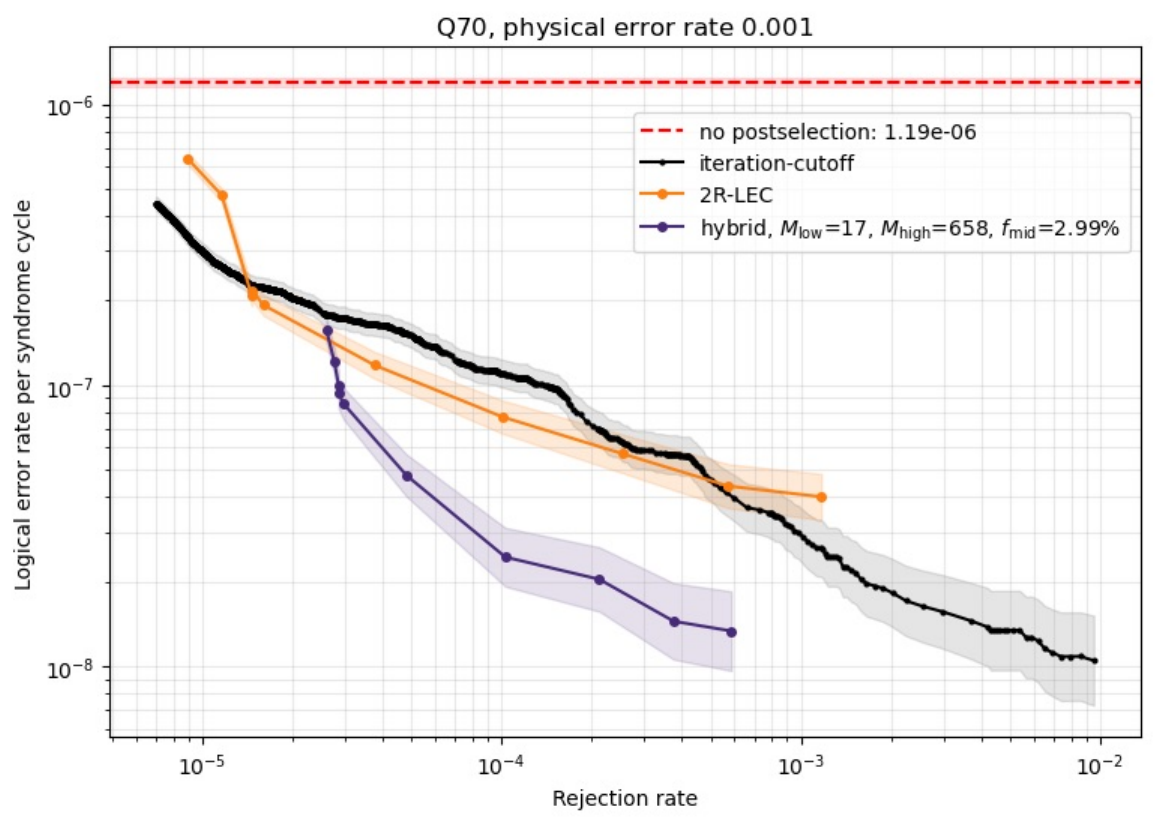}
    }
    \hfill
    \subfloat{
        \includegraphics[width=0.48\textwidth]{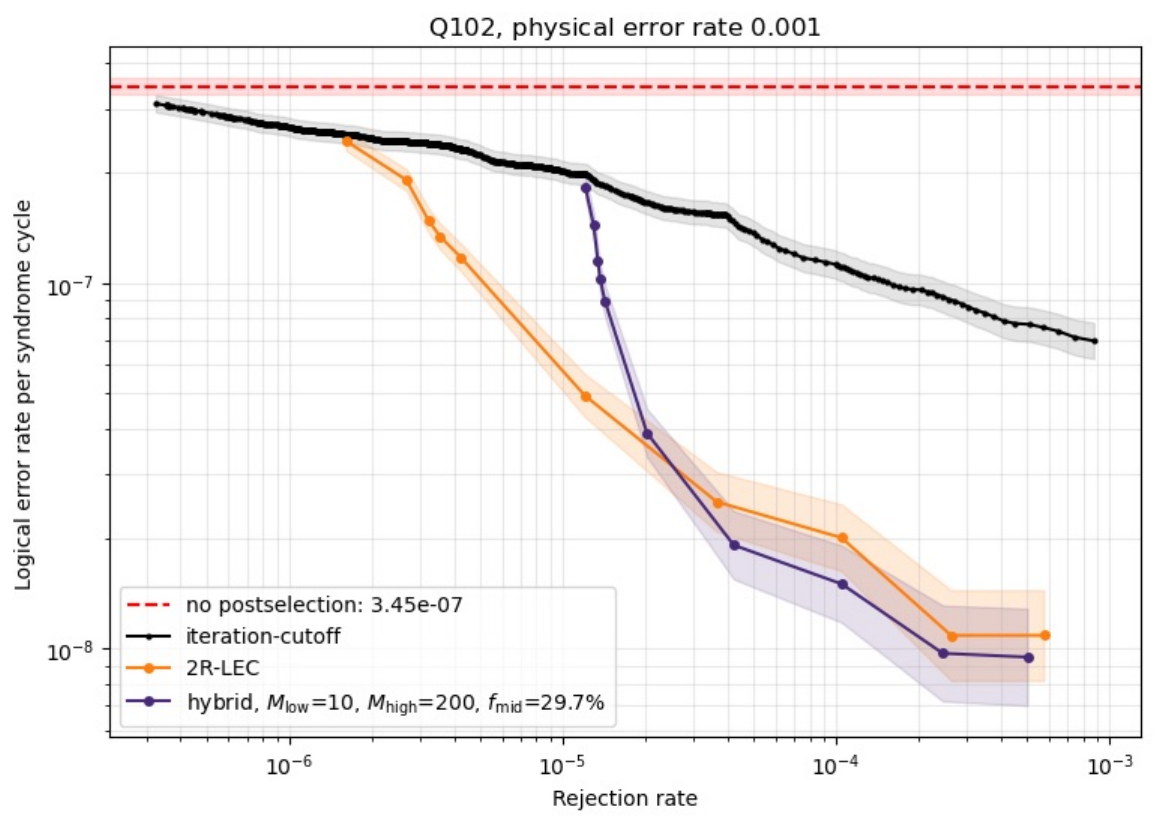}
    }
    \caption{Comparison of the iteration-cutoff rule, the hybrid rule, and 2R-LEC at $p=10^{-3}$ for (left) Q70 and (right) Q102. The thresholds $\Mlow$ and $\Mhigh$ and the fraction $f_{\mathrm{mid}}$ are defined in Sec.~\ref{sect:soft_info}.}
    \label{fig:hybrid_global}
\end{figure*}

\section{Circuit-level simulation results}
\label{sect:simulation_results}

We present circuit-level simulation results for the five code instances listed in Sec.~\ref{sect:soft_info}. For BB90 and BB144, we simulate physical error rates of $p=10^{-3}$ and $p=3\times10^{-3}$ using the circuits and circuit-level noise model from~\cite{bravyi2024high}. For Q54, Q70, and Q102, we simulate $p=10^{-3}$ using the circuits and circuit-level noise model from~\cite{tripier2026fault}. Unless stated otherwise, all simulations use memory-$X$ circuits and the \texttt{beam8\_230iters} configuration of the beam search decoder described in Sec.~\ref{sect:soft_info}. Section~\ref{sect:experimental_data} uses a different beam search configuration.

Throughout this paper, shaded bands in the plots denote $95\%$ Wilson score confidence intervals~\cite{wilson1927probable} for the logical error rate per syndrome cycle, computed from the binomial statistics of the accepted shots at each cutoff.

\subsection{Soft global decoder}

We first compare iteration-cutoff post-selection with the no-post-selection baseline for BB90 and BB144. Figure~\ref{fig:BB_90_144} shows the results at $p=10^{-3}$ and $3\times10^{-3}$. These simulations use the global beam search decoder, with the number of syndrome extraction cycles set equal to the code distance.

At $p=3\times 10^{-3}$, rejecting $1\%$ of shots suppresses the logical error rate by factors of $7.4\times$ and $22.6\times$ for BB90 and BB144, respectively. At $p=10^{-3}$, rejecting only $0.1\%$ of shots yields suppression factors of $124\times$ and $78\times$, respectively.

Figure~\ref{fig:hybrid_global} compares the iteration-cutoff rule, the hybrid rule, and 2R-LEC for Q70 and Q102 at $p=10^{-3}$.

For Q70, at a rejection rate of $10^{-4}$, the iteration-cutoff, hybrid, and 2R-LEC rules achieve logical-error suppression factors of $10.8\times$, $48.9\times$, and $15.5\times$, respectively. The hybrid rule has $f_{\mathrm{mid}}<3\%$, so fewer than $3\%$ of shots require a second decoder pass. 
For Q102, at the same rejection rate, the three rules achieve suppression factors of $3.1\times$, $23.0\times$, and $17.1\times$, respectively. 


\begin{table}
\centering
\caption{Soft decoder parameters used in Table~\ref{tab:global_soft_comparison}. $\Mcut$ is used for iteration-cutoff. The same $b$ is used for both 2R-LEC and hybrid.}
\label{tab:global_soft_parameters}
\small
\setlength{\tabcolsep}{4pt}
\renewcommand{\arraystretch}{1.0}
\begin{tabular*}{\columnwidth}{@{\extracolsep{\fill}}lcccc@{}}
\toprule
Code & $\Mcut$ & $b$ & $\Mlow$ & $\Mhigh$ \\
\midrule
BB90  & $496$ & $1.04$ & $85$  & $653$ \\
BB144 & $139$ & $1.31$ & $90$  & $136$ \\
\midrule
Q70   & $261$ & $1.20$ & $17$  & $658$ \\
Q102  & $90$  & $1.30$ & $10$  & $200$ \\
Q54   & $172$ & $1.25$ & $150$ & $173$ \\
\bottomrule
\end{tabular*}
\end{table}

Tables~\ref{tab:global_soft_comparison} and~\ref{tab:global_soft_parameters} summarize the soft decoder's performance and selected parameters at $p=10^{-3}$. When tuning parameters, we first select the 2R-LEC reweighting parameter $b$, defined in Eq.~\eqref{eq:2R-LEC}, using a calibration sweep. We then hold $b$ fixed and choose $(\Mlow,\Mhigh)$ to maximize hybrid error suppression subject to the rejection rate constraint. The iteration cutoff $\Mcut$ is chosen separately to match the target rejection rate. For Q54, we choose a target rejection rate of $10^{-3}$ rather than $10^{-4}$ because rejecting non-converged shots alone imposes a rejection-rate floor of approximately $1.24\times10^{-4}$ for the iteration-cutoff rule. The ``Walking cat" noise model in Table~\ref{tab:global_soft_comparison} specifically refers to the ionic circuit-level noise model introduced in \cite{tripier2026fault}.

At these operating points, hybrid soft decoding gives the largest logical error suppression for every code while using approximately $32\%$--$65\%$ of the mean BP iterations required by 2R-LEC. Iteration cutoff remains the least costly rule; for BB90, BB144, and Q54, hybrid improves suppression with less than $3\%$ additional BP work relative to cutoff. Mean BP iterations are averaged over all attempted shots, including rejected shots.

\subsection{Soft sliding-window decoding of logical measurements}

Using the sliding-window extension of the iteration-cutoff rule described in Sec.~\ref{sect:soft_info}, we decode EDM-4 memory-$Z$ circuits for Q54 and Q70. The sliding-window decoder uses window size $w=5$ and commit size $c=3$, with the beam search decoder serving as its inner decoder.

An $r$-round error-detected measurement (EDM-$r$) performs $r$ measurements of a logical Pauli operator $\bar{P}$ using cat states. These \emph{cat-based measurements}, defined in Sec.~XIII of~\cite{tripier2026fault}, yield error-corrected outcome bits $y_1,\ldots,y_r$ after decoding. The protocol returns the common value $\hat{y}=y_1=\cdots=y_r$ only if all $r$ outcomes agree; otherwise, it aborts and the shot must be restarted. Following~\cite{tripier2026fault}, we refer to the probability that EDM returns an incorrect value as the logical bit error rate.

We adopt the simulation circuit of~\cite{tripier2026fault}. An EDM-4 circuit begins and ends with $d$ syndrome extraction cycles (SECs), where $d$ is the code distance, and consecutive cat-based measurements (CMs) are separated by one SEC. The circuit therefore contains four CMs and $2d+3$ SECs in total. Following~\cite{ye2026real}, we group each CM with the immediately following SEC and call the resulting unit a \emph{cat SEC}. The circuit can then be viewed as $d$ ordinary SECs, followed by four cat SECs and $d-1$ ordinary SECs.

With $(w,c)=(5,3)$, each decoding window except possibly the final one contains five ordinary or cat SECs. Importantly, a cat SEC has the same Tanner graph as an ordinary SEC; the two differ only in the prior probabilities assigned to certain error variables~\cite{ye2026real}. The decoder can therefore process all windows, apart from possible boundary-window exceptions, using the same message-passing graph, updating only the relevant priors when a window contains one or more cat SECs.

The decoder uses the syndrome data to correct the four CM readouts. In these memory-$Z$ circuits, all data qubits are initialized to $|0\rangle$, and each CM measures the same logical $Z$ operator. With the $+1$ eigenvalue represented by the binary value $0$, the correct decoded outcome of each CM is therefore $0$. Thus, the corrected pattern \texttt{0000} is accepted as a successful outcome, whereas \texttt{1111} is accepted by EDM but counted as a logical bit error. Any mixed four-bit pattern fails the EDM consistency check and is rejected regardless of the iteration cutoff from the soft decoder.

\begin{figure}
    \centering
    \includegraphics[width=\linewidth]{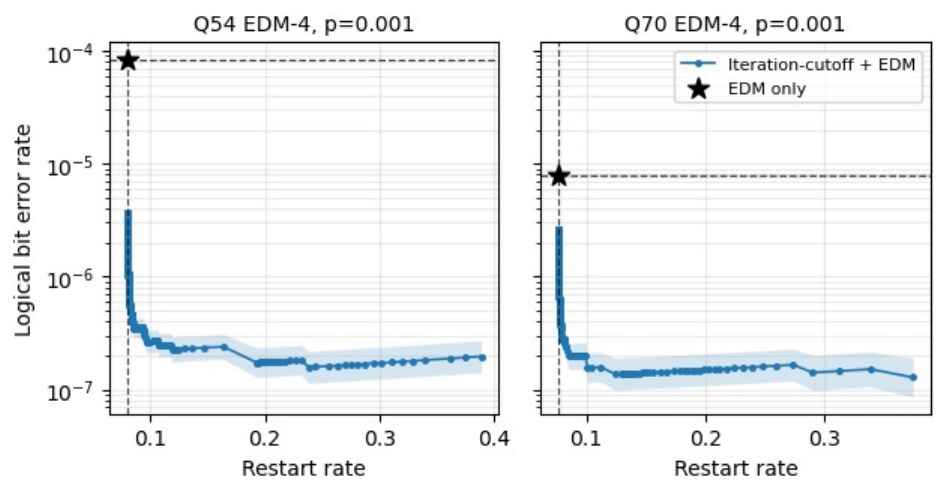}
    \caption{Logical bit error rate versus restart rate for the $(w=5,c=3)$ soft sliding-window decoder applied to EDM-4 at $p=10^{-3}$ for (left) Q54 and (right) Q70. Each curve plots the tradeoff obtained by varying the iteration cutoff $\Mcut$.}
    \label{fig:soft_EDM4}
\end{figure}

We define the restart rate as the probability that a shot is rejected by the EDM consistency check, by the soft decoder, or by both. Shots rejected by both criteria are counted only once. Since each rejection requires a restart, the restart rate equals the rejection rate; we use the former term to follow the EDM terminology in~\cite{tripier2026fault}. Figure~\ref{fig:soft_EDM4} plots the logical bit error rate against restart rate at $p=10^{-3}$ as $\Mcut$ is varied along the curves.

For Q54, increasing the restart rate from the $7.9\%$ EDM-only baseline to $8.4\%$, suppresses the logical bit error rate by a factor of $210\times$. For Q70, increasing the restart rate from $7.7\%$ to $8.2\%$ yields a suppression factor of $30\times$.

In the walking cat architecture, the MEK and CH2 factories use EDMs to prepare magic states encoded in Q70 and Q54, respectively~\cite{tripier2026fault}. Our results suggest that applying soft beam search decoding to these EDMs could reduce the logical error rates of these factories. The soft decoder can also reduce the logical error rate of Clifford gates implemented through the recently proposed logical CliNR scheme, which relies on a resource state prepared using EDMs~\cite{webster2026fast}.

\section{Beyond-breakeven trapped-ion memories through post-selection}
\label{sect:experimental_data}

We apply the iteration-cutoff rule directly to experimental qLDPC memory data collected on an IonQ trapped-ion quantum computer and reported in~\cite{tham2026breakeven}. The data comprise $X$- and $Z$-basis memory experiments with one to six syndrome cycles and both prepared logical eigenstates in each basis. We consider three instances of the BB5 code family~\cite{ye2025quantum}, namely $[[18,4,3]]$, $[[24,4,4]]$, and $[[30,4,5]]$ codes, as well as the GB4 $[[16,2,4]]$ and $[[26,2,5]]$ codes~\cite{kovalev2013quantum}. We re-decode the recorded detector outcomes using error priors derived from device calibration; no additional quantum experiments are required.

For a direct comparison with Ref.~\cite{tham2026breakeven}, we decode every experimental shot using the same \texttt{beam32\_340iters} configuration of the beam search decoder~\cite{ye2026beam} employed in that work. For each shot, we record both the correction and its convergence iteration $M(\boldsymbol{s})$. The maximum possible convergence iteration for this configuration is $M_{\max}=11500$. Every shot in the experimental dataset converges, and the largest observed value of $M(\boldsymbol{s})$ is $11112$. Consequently, setting $\Mcut=11500$ accepts every shot and defines the exact hard-decoding, no-post-selection reference.

\begin{figure*}[t]
    \centering
    \subfloat{
        \includegraphics[width=0.48\textwidth]{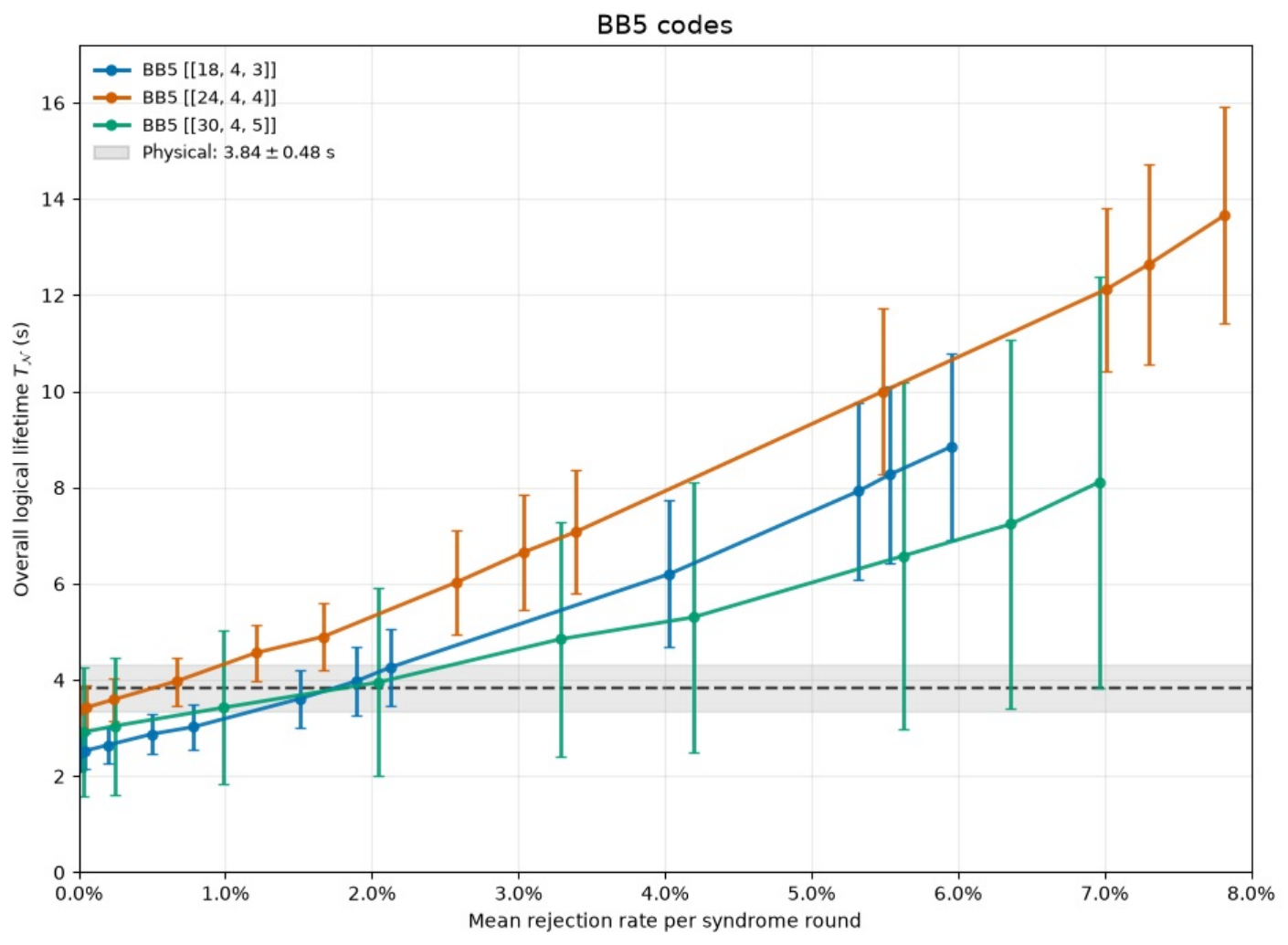}
    }
    \hfill
    \subfloat{
        \includegraphics[width=0.48\textwidth]{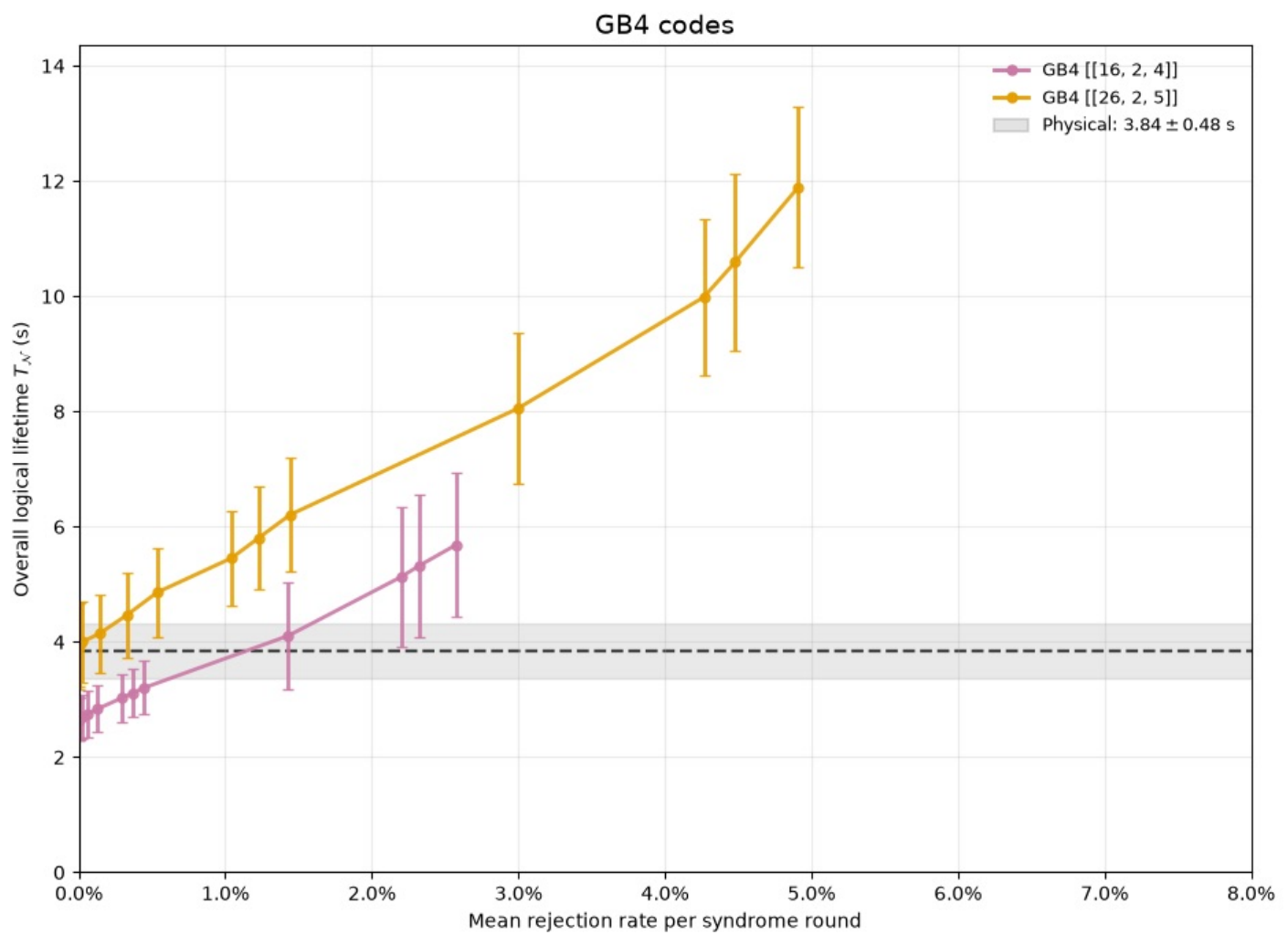}
    }
    \caption{Overall logical lifetime $T_{\mathcal N}$ versus mean rejection rate per syndrome round under iteration-cutoff post-selection for the experimental (left) BB5 and (right) GB4 memory data. Each point corresponds to a different value of $\Mcut$. Vertical error bars denote $95\%$ confidence intervals from the lifetime fits. The dashed line and gray band mark the measured physical-qubit lifetime, $3.84\pm 0.48\,\mathrm{s}$.}
    \label{fig:experimental_lifetime}
\end{figure*}

For a chosen cutoff $\Mcut$, we retain only shots satisfying
$M(\boldsymbol{s})\leq\Mcut$. Varying $\Mcut$ produces the lifetime--rejection tradeoff curves shown in Fig.~\ref{fig:experimental_lifetime}. Since the experiments contain different numbers of syndrome cycles, we report the mean rejection rate per syndrome round.

At each cutoff, we pool the two prepared logical eigenstates within each basis. Let $\epsilon_{\mathrm L}(N_{\mathrm{cyc}})$ denote the probability, conditioned on acceptance, that at least one of the $k$ logical qubits is decoded incorrectly after $N_{\mathrm{cyc}}$ syndrome cycles. We fit the corresponding logical survival probability to
$$
1-\epsilon_{\mathrm L}(N_{\mathrm{cyc}})
= C\left( \frac{1+e^{-N_{\mathrm{cyc}}/\tau}}{2} \right)^k,
$$
where $C$ and $\tau$ are fitting parameters. We perform the survival-probability fitting separately for the $X$- and $Z$-basis data, obtaining decay constants $\tau_X$ and $\tau_Z$ in units of syndrome cycles. Let $T_{\mathrm{cycle}}$ denote the code-dependent duration of one syndrome cycle. The corresponding basis-specific logical lifetimes are
$$
    T_X=\tau_X T_{\mathrm{cycle}}, \qquad
    T_Z=\tau_Z T_{\mathrm{cycle}}.
$$
Following the effective Pauli-channel model of~\cite{tham2026breakeven}, the overall logical lifetime is then defined by
$$
    \frac{1}{T_{\mathcal N}}
    = \frac{1}{3} \left( \frac{1}{T_X}+\frac{1}{2T_Z} \right).
$$
We compare $T_{\mathcal N}$ with the measured physical-qubit lifetime $T_{\mathcal N}^{\mathrm{phys}}=3.84\pm 0.48\,\mathrm{s}$. An encoded memory achieves beyond-breakeven performance when its logical lifetime exceeds this physical-qubit lifetime.

Figure~\ref{fig:experimental_lifetime} shows that the fitted logical lifetime increases monotonically with the rejection rate for all five codes. At zero rejection, the logical lifetimes remain near the physical-qubit reference, consistent with the breakeven performance obtained using hard decoding in~\cite{tham2026breakeven}. Rejecting only $2.6\%$--$5.6\%$ of shots per syndrome round is sufficient to more than double the logical lifetime of every code. At these modest rejection rates, the central estimates of all five logical lifetimes exceed the physical-qubit lifetime. For four of the five codes, the non-overlapping logical- and physical-lifetime uncertainty intervals demonstrate statistically significant beyond-breakeven performance; for the BB5 $[[30,4,5]]$ code, the uncertainty intervals still overlap.

Tighter cutoffs yield still larger lifetime gains. The BB5 $[[24,4,4]]$ code reaches $13.66\,\mathrm{s}$ at a rejection rate of $7.82\%$, approximately four times both its no-post-selection value and the physical-qubit lifetime. The GB4 $[[26,2,5]]$ code similarly reaches $11.90\,\mathrm{s}$ at a rejection rate of $4.91\%$, approximately three times its no-post-selection value. These results show that further lifetime gains remain available when a larger rejection budget is acceptable.

\section{Conclusion}

We proposed a soft version of the beam search decoder for quantum LDPC codes and demonstrated that it improves the logical error rate of quantum error-correcting codes at the price of a small rejection rate.
The strength of our approach is that it leads to a post-selection strategy based on internal decoder data and it does not require additional computation.
Moreover, it consumes fewer BP iterations than previous approaches such as the 2R-LEC method~\cite{xie2026simple} or the forced-gap method~\cite{wills2026forced}.

This work already applies to current QEC experiments \cite{tham2026breakeven}, allowing to reach beyond-breakeven performance for LDPC codes.
Moreover, our soft decoder can be immediately integrated in the fault-tolerant quantum computing architectures based on LDPC codes to improve the logical error rate of logical operations performed through resource states prepared offline at the price of a small increase in the restart rate.

In the walking cat architecture, our simulations of EDMs show that the soft beam search decoder can be used to reduce the logical error rate of magic factories~\cite{tripier2026fault} and fast logical Clifford gates~\cite{webster2026fast}, which translates into an increased success probability for concrete applications such as solving the elliptic-curve discrete logarithm problem~\cite{haner2026computing}.

\section*{Acknowledgment}

The authors thank John Gamble and the Quantum Error Correction team at IonQ for their insightful feedback.

\bibliography{references}

\end{document}